# Supporting Answers with Feedback in Social Q&A

**John Frens**
University of Washington
Seattle, WA
jfrens@uw.edu

**Erin Walker**
Arizona State University
Tempe, AZ
eawalke@asu.edu

**Gary Hsieh**
University of Washington
Seattle, WA
garyhs@uw.edu

**ABSTRACT**
Prior research has examined the use of Social Question and Answer (Q&A) websites for answer and help seeking. However, the potential for these websites to support domain learning has not yet been realized. Helping users write effective answers can be beneficial for subject area learning for both answerers and the recipients of answers. In this study, we examine the utility of crowdsourced, criteria-based feedback for answerers on a student-centered Q&A website, Brainly.com. In an experiment with 55 users, we compared perceptions of the current rating system against two feedback designs with explicit criteria (Appropriate, Understandable, and Generalizable). Contrary to our hypotheses, answerers disagreed with and rejected the criteria-based feedback. Although the criteria aligned with answerers' goals, and crowdsourced ratings were found to be objectively accurate, the norms and expectations for answers on Brainly conflicted with our design. We conclude with implications for the design of feedback in social Q&A.

**INTRODUCTION**
Social Question and Answer (Q&A) websites offer a space where users come together at scale to network, exchange information, and learn [16]. Whether from schools or massive online courses, students find Q&A sites while searching the internet for help with assignments or course material [15]. Each month, 80 million users land on Brainly.com, a site which serves middle school, high school, and college level Q&A [22].

Although users participate in Q&A for learning purposes [4], not all, or even most, of answers are high-quality [1], which is a problem for students seeking to learn. We define *high-quality* answers in this context as those that meet the needs of readers and provide sufficient detail to be understood and internalized. Figure 1 exemplifies a high-quality question and answer exchange from Brainly. Previous study of Brainly [11] reported high variance in the readability of answers, suggesting that many are so poorly written that they are difficult to understand, and a deletion rate as high as 30%, demonstrating that community moderators find a significant proportion of answers to be inaccurate or irrelevant.

**Figure 1: An answer on Brainly.com. Feedback is given to answerers by clicking the 'thanks' button, rating the answer from 1 to 5 stars, or adding a comment.**

There is a huge opportunity for Q&A to support informal, self-directed learning by encouraging high-quality answers. For *Answerers*, the users who write answers, the process of providing quality help involves using domain knowledge, monitoring accuracy, and engaging in explanation and elaboration [14]. These are self-regulated learning processes recognized to promote robust learning in more traditional collaborative learning scenarios [21]. Learning through answering is also one of the most prominent motivations reported by Brainly answerers [4]. Moreover, an accurate explanation that addresses the question's misconceptions will better support the needs of *askers*, the users who post questions [3,19]. And *viewers* of the Q&A website's searchable archive may better be able to make use of high-quality answers [15]. Encouraging high-quality answers is crucial for facilitating learning during the information exchange on social Q&A. Thus, we ask: How can we encourage high-quality answers in social Q&A?

Giving answerers feedback may be a valuable mechanism for encouraging high-quality answers. Feedback is a common strategy for improving the quality of student work, and studies have shown that the appropriate use of feedback is extremely powerful [8]. Feedback affordances in social Q&A could support answerers as they learn to give effective help by providing directions for improvement.

However, feedback is currently limited on Q&A sites in both presence and substance. As we will show, almost half of answers on Brainly do not receive any feedback, and 1- to 5-star ratings, the most common form of feedback exchanged on Brainly, are likely insufficient for encouraging high-quality answers. This greatly undermines potentially valuable opportunities for answerers to improve.

In a mixed methods study, we explored the use of crowdworkers and tested the effectiveness of a criteria-based feedback intervention on Brainly answerers. Crowdworkers are a promising source of scalable feedback for MOOCs and other learning settings, and have shown the capability to give expert-level feedback when given criteria [20].

We gathered 1- to 5-star ratings from crowdworkers under three different conditions. In the **no-criteria** condition, crowdworkers simply rated the quality of the answer without criteria. In two experimental conditions, we asked crowdworkers to rate on three feedback criteria based on Webb [19]: Appropriate, Understandable, and Generalizable. The **unidimensional** condition was a single rating for the criteria, while the **multidimensional** condition had multiple ratings—one per criterion. We hypothesized:

> H1: Participants in criteria-based conditions will perceive feedback as more useful and be more likely to improve their answers, compared to those in the no-criteria condition.

We further explore whether the increased detail of multiple ratings, one per criterion, are more useful. Hattie and Timperley [8] state feedback should direct students about how to proceed. Three ratings, one for each criterion, may better indicate where to focus efforts. Thus, we hypothesize:

> H2: Participants in the multidimensional condition will perceive feedback as more useful and be more likely to improve their answers than those in the unidimensional condition.

Contrary to our expectations, Brainly answerers in the no-criteria condition were more likely to agree with the feedback, find it helpful, and state it will be easy to incorporate. To explore this surprising result, we interviewed participants about their motivations, perception of the experimental criteria and usage of feedback on Brainly. While answerers agreed with the criteria in principle, they placed more importance on meeting the expectations of the asker. However, relying on askers to set the standard for answerers may ultimately inhibit domain learning of askers, answerers, and viewers during use of social Q&A. This study makes several important contributions:

- Our experimental results expose critical challenges with applying classroom theory in the context of social Q&A
- Our interview results contextualize feedback within Q&A, revealing how an asker-focused outlook affects answerers' goals and feedback expectations
- We present design recommendations based on our findings for feedback within social Q&A, with the goal of promoting high-quality answers that can support the self-regulated learning of answerers, askers, and viewers.

## RELATED WORKS
### Social Q&A on Brainly

Social Q&A services create a participation-based community for users to ask questions and others to answer them [16]. Study of Q&A has primarily focused on predicting the quality of questions and answers [1,2,11] and understanding user motivations [3,4,6]. Supporting students within social Q&A by integrating learning is a research area that holds great potential for educational outcomes [16].

Brainly.com, the site under study, is a social Q&A website that combines social networking elements with academic Q&A for students in middle school, high school, and college. Learning is an important goal for users on Brainly, whether it's gaining knowledge in a favorite subject, verifying information, or learning through answering questions [4]. Helpfulness, informativeness, and relevance of answers are considered by this community to be most important [2].

Answer quality considerably varies [11], and with only up to two answers afforded per question, many exchanges go without a high-quality answer. Thus, there is a need for answerers to learn to give more effective help. The central motivation of our study is to support answerers in learning to provide quality help to their peers, which, in turn, we believe will contribute to domain learning outcomes for all users.

### Encouraging High-Quality Answers with Feedback

There is great potential for feedback to teach answerers to give effective help. Feedback is the provision of information regarding task performance. A meta-analysis of feedback in classrooms suggests it can yield impressive performance gains given the right conditions [9]. To be effective, feedback should support well-defined goals, be related to achieving success on critical dimensions of the goal, and consist of information about how to progress or proceed [8]. It should address the accuracy of a learner's response and may touch on particular errors and misconceptions [17]. Applying these principles in social Q&A settings by making goals explicit and increasing the specificity of feedback may be greatly beneficial for answer quality.

The feedback currently available on Brainly is limited in its capacity to address specific goals and provide direction to answerers. Feedback mechanisms include comments, star-ratings, the 'thanks' button, and the 'Brainliest answer' tag– a marker indicating the best answer. Although a high-quality answer may receive a five-star rating, 'thanks,' or 'Brainliest answer,' these affordances do not explicitly encourage quality. Comments, a good candidate for providing direction, are sparse, occurring on only 20% of our answer sample. Thus, there is an unmet need on Brainly for feedback.

**Crowd Feedback in Learning Settings**

Crowdsourcing techniques offer a promising solution for providing feedback in large-scale learning settings, especially where automated grading is infeasible. For instance, Fraser et al presented CritiqueKit, a system that reduces the burden of providing feedback at scale by classifying feedback and providing recommendations to reviewers [7]. Kulkarni et al built PeerStudio, a platform for rubric-based peer assessment on open-ended, in-progress student work, and demonstrated that rapid feedback given through this system helped to improve overall quality of work [10]. These systems have been shown to be effective for reducing burden and increasing quality in peer review, and have great potential to improve feedback in informal spaces. However, they rely on motivating groups of peers to review, which poses an open challenge in social Q&A.

In this study, we explore the use of crowdworkers for providing feedback in social Q&A. Given appropriate support, crowdworkers are effective at giving feedback on student work. Luther et al [12], in a study of the crowdworker-sourced critique system CrowdCrit, found that aggregated crowd critique approached expert critique, and designers who received crowd feedback perceived that it improved their design process. In a study of design feedback by Yuan et al [20], crowdworkers providing feedback without criteria tended to focus on surface-level features, while with criteria, their reviews were as valuable as expert reviews. In our study, we hope to effectively deploy crowd feedback in a setting where peer feedback is rare, through the use of aggregated crowdworker feedback with criteria targeted toward promoting effective help-giving.

**Supporting Domain Learning within Social Q&A**

Classroom studies of peer tutoring provide insights for supporting domain learning within social Q&A. In classroom settings, the positive effects of peer tutoring on domain learning are documented across subjects and grade levels [5]. The beneficial effects are as significant for the tutor and as they are for the tutee, a phenomenon known as the *tutor learning effect* [14]. Tutor learning occurs due to the intermingled processes of *metacognition*, defined as self-monitoring, understanding, and recognizing misconceptions, and *knowledge construction*, the repair of knowledge gaps, elaboration of knowledge, and generation of new ideas [14].

These processes are invoked as students give explanatory, step-by-step answers [14]. Thus, encouraging high quality also supports answerers' goal of learning through answering.

Study of peer help has uncovered key dimensions for the helpfulness of answers. Based on several classroom studies, Webb [19] enumerates conditions necessary for peer help to benefit the recipient's learning. First, the help must be relevant, correct, and complete. We refer to this as *Appropriate* in our work. Second, the student receiving the help must understand the explanation, which we refer to as *Understandable*. And third, which we call *Generalizable*, the student must be able to internalize the help. This is demonstrated when the student subsequently solves another problem. Timing is also important, as students may forget context if the help is given too late. In Q&A, the answerer has little control over the timing of the help or the reader's subsequent behavior.

Classroom peer help differs from online Q&A in important ways, for instance, asynchronous communication and public archival. However, a translation of Webb's findings could help incorporate learning into the space of information exchange on social Q&A. Based on Webb's framework, we built a set of three criteria to assess the effectiveness of answers given over social Q&A (Table 1).

| Criterion | Definition |
| --- | --- |
| Appropriate | The asker's need is addressed by the answer<br>▪ Relevant to the asker's need<br>▪ Accurately states the correct answer<br>▪ Completely addresses the asker's need |
| Understandable | The asker can understand the answer<br>▪ Well-formulated, grammatically correct, and clear<br>▪ Accounts for what the asker knows or doesn't know<br>▪ Translates difficult vocabulary into familiar terms |
| Generalizable | The asker can use the answer to solve similar problems<br>▪ Explains a strategy for finding the answer<br>▪ Or, provides an example<br>▪ Or, links to resources showing where the answer was found |

**Table 1: Description of the answer evaluation criteria used in the experimental conditions of the study.**

These criteria introduce specific goals to be evaluated using 1- to 5-star rating feedback. Kluger and DeNisi [9] state the motivation to respond to feedback is driven by feedback-standard discrepancies. Therefore, we expect that a 1- to 5-star rating presented alongside criteria with specific dimensions should be more effective than a 1- to 5-star rating presented without criteria.

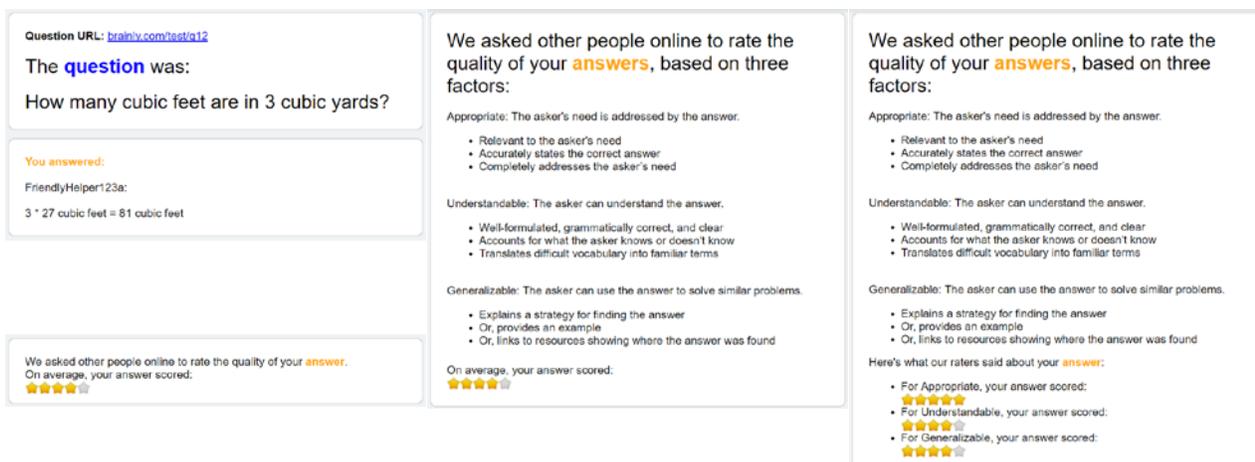

**Figure 2: Q&A pair (top left) with no-criteria (bottom left), unidimensional (middle), and multidimensional (right) feedback conditions. The Q&A pair was shown together with each rating.**

## FEEDBACK INTERVENTION STUDY
### Community
Brainly.com is a community and website with affordances for Q&A and social networking. The questions asked on Brainly span 18 academic topics including mathematics, history, English, biology, chemistry, and physics. Students ask questions on Brainly to receive immediate homework help, gain knowledge in domains of interest, and verify existing knowledge [4]. The site was chosen for the study due to its popularity and impact: Brainly serves 80 million unique users monthly [22].

### Participants
We recruited answerers over the age of 18 who answered at least 5 questions within the past month. Brainly's community managers sent out a message to 2,000 users: "Interested in winning a [company] gift card? We are conducting a new Brainly research study with University of Washington, to help develop a better online learning and need your help. If you would like an opportunity to win (or just help us out!) please click the link below to provide your email address and we will reach out to you soon!" 103 answerers responded to our initial survey. 87 of those respondents were recruited into the experiment and had their answers evaluated by crowdworkers. 16 did not meet our requirements because of incomplete responses, being under the age of 18, or providing an incorrect/absent Brainly username. During the experiment, another 32 users who did not view the feedback were dropped. 55 Brainly users completed the study. They were primarily college aged, with a mean age of 24 and median age of 21. Their self-reported ethnicities were: 22 Caucasian, 14 Asian, 10 Hispanic, 2 Black, 1 Native American, 3 other, and 3 unreported. Their self-reported genders were: 32 male, 22 female, and 1 unreported.

### Procedure
We designed an experiment to compare three feedback designs: no-criteria, unidimensional, and multidimensional. We collected Brainly users' five most recent answers and gathered crowdworker ratings. Mechanical Turk raters were randomly assigned to one of the three conditions. They rated up to 20 answers, all written by different users, and were compensated 5 cents per each rating, which took a median 24 seconds. Each answer was rated by three crowdworkers under each of the three conditions, thus each answer was rated 9 times. Under the no-criteria condition, crowdworkers simply rated answers quality from 1 to 5 stars, and Brainly users viewed the average of 3 crowdworker ratings for each answer. Under the unidimensional condition, crowdworkers read the criteria and rated the quality of answers from 1 to 5 stars. Brainly users viewed the criteria and average ratings. For the multidimensional condition, crowdworkers read the criteria and provided separate 1- to 5-star ratings for each criterion. Brainly users viewed the criteria and average ratings for each criterion. Figure 2 demonstrates feedback under each condition.

We randomly assigned Brainly users to view one of the three feedback conditions, a between-subjects design. We sent each of the users an email with a link to their individualized feedback webpage, which showed the mean crowdworker ratings for each of the five answers. In the unidimensional and multidimensional conditions, the criteria were shown at the top of the feedback webpage. Brainly users gave 5-point Likert-scale responses to the following four statements after viewing rating: "I agree with the feedback", "I found the feedback helpful," "I will change my answers in the future, based on the feedback," and "It will be easy to incorporate the feedback into my future answers." To measure behavioral outcomes, we gathered archival data on the 5 answers written by users before and after the experiment. We computed average community ratings, summed the 'thanks' received, and counted the number of 'Brainliest' answers.

### Analyses
To assess the difference in crowdworker ratings between rating conditions, we used a mixed linear regression. As crowdworker ratings were the aggregate of three ratings, the dependent variable was approximately continuous. We tested whether the presence of criteria significantly affected the rating that crowdworkers gave. Condition was the single fixed effect, while answer was included as a random effect. The dependent variable was the average rating of three crowdworkers, from 1 to 5 stars.

To examine H1 and H2, we tested the effect of the feedback conditions on the 5-point Likert measures using a series of four logistic regressions. Likert scales are not interval, i.e. the differences between each level are not equal. We split the measures into two categories, with "agree and strongly agree" in the positive category and "neutral," "disagree," and "strongly disagree" in the negative category. Condition and rating were included in the regression model as fixed effects, while user was a random effect. The dependent variable was the likelihood of selecting "agree" or "strongly agree.

To assess behavioral outcomes, we examined the pre-post differences in community feedback on participants' answers. For each user, we subtracted the mean rating of 5 answers written before the experiment from the mean rating of 5 answers written after the experiment. The same was done for the sum number of 'thanks', and the number of answers marked 'Brainliest answer.' We used an ANOVA to test whether users' outcomes differed between conditions, where the pre-post difference was the dependent variable, and condition was the independent variable.

**Results**
*Answer Dataset*
The answer dataset consisted of the five most recent answers from 55 participants. One participant deleted 2 of their answers during the experiment, and as a result our dataset includes 273 answers. To contextualize the data, we present information about the feedback each answer received on Brainly.com. Of the 273 answers, 6 are omitted here because due to deletion by users or community moderators before the time of the web scraping. The results are shown in Table 2.

| Feedback Received | Number of Answers | Percentage of Answers |
|---|---|---|
| Brainliest | 22 | 8.2% |
| Comment | 54 | 20.2% |
| Rating | 116 | 43.4% |
| Thanks | 102 | 38.2% |
| No Feedback | 114 | 42.7% |

**Table 2: The type of feedback received by 267 answers, along with the number and proportion of answers receiving that type of feedback.**

A small majority of answers in our dataset received feedback on Brainly, with the most common types being ratings and 'thanks'. Ratings by Brainly users were generally positive: 87.1% 5-star, 6.0% 4-star, 1.7% 3-star, 0% 2-star, and 5.2% 1-star. A relatively small proportion of answers received 'Brainliest answer' or comments. A noteworthy 42.7% of answers in our sample did not receive feedback of any kind.

*Agreement and Accuracy of Crowdsourced Ratings*
We next examine the difference in crowdsourced ratings between conditions, and relationships between ratings from crowdworkers, Brainy users and experts. Crowdworkers provided 2,457 ratings for the 273 answers—each answer was rated by 3 crowdworkers per condition. Very high levels of agreement between crowdworkers would indicate that multiple ratings are redundant and possibly unnecessary, while low levels of agreement would indicate that crowdworkers provide an unreliable assessment of these constructs. As seen in Table 3, Cronbach's α for the ratings ranged from 0.47 to 0.59. These values indicate a moderate level of interrater agreement between crowdworkers about the quality of answers. The finding suggests that aggregating over multiple ratings is likely to be important for generating a valid rating, but that the crowdworkers trended together.

| Condition | Mean | SD | α |
|---|---|---|---|
| No-criteria | 3.98 | 0.88 | 0.49 |
| Unidimensional | 3.78 | 0.92 | 0.57 |
| Multidimensional - Averaged | 3.89 | 0.81 | |
| Appropriate | 4.19 | 0.87 | 0.59 |
| Understandable | 4.06 | 0.83 | 0.47 |
| Generalizable | 3.44 | 1.11 | 0.59 |

**Table 3: Mean, standard deviation and Cronbach's α for 1-to-5 star crowdworker ratings of 273 answers, by condition, with averaged and individual criterion shown for the multidimensional condition.**

As shown in Table 3, the mean rating given to answers under the no-criteria condition was higher than the unidimensional or multidimensional conditions. A mixed linear regression confirms that unidimensional ratings were significantly lower than no-criteria ratings (β=-0.22, p<0.001), and the multidimensional ratings trended lower with marginal significance (β=-0.10, p=0.09). Within the multidimensional condition, the mean Generalizable rating was lower than Appropriate or Understandable. A mixed linear regression (with criterion as the fixed effect and answer as the random effect) reveals that Appropriate was significantly higher than Understandable (β=0.14, p<0.01), and Generalizable was significantly lower than Understandable (β=-0.72, p<0.001). These differences between no-criteria and unidimensional ratings may suggest that raters were more critical when provided with criteria for the answers. Generalizable was the lowest rated category in the multidimensional condition, suggesting low generalizability in the answers was the main reason for lower ratings when the criteria were present.

Differences between no-criteria crowdworker ratings and Brainly user ratings could indicate that crowdworkers have a different mental model for rating answers than Brainly users. We compared crowdworker ratings with Brainly user ratings using the subset of 116 answers that received ratings on Brainly. We find a weak Pearson correlation of 0.23 (p<0.01) between Brainly user ratings and crowdsourced ratings in the no-criteria condition. This suggests that Brainly users view the 5-star system differently from crowdworkers.

| Criterion | Expert Rating | Crowd Rating |
|---|---|---|
| Appropriate | 4.25±0.97 | 4.17±0.81 |
| Understandable | 4.16±0.76 | 4.02±0.82 |
| Generalizable | 2.88±1.33 | 3.39±1.13 |

**Table 4: Mean ± Standard Deviation of crowdworker and expert ratings for a 50 answer sample.**

If crowdworker ratings are not sufficiently representative of Appropriate, Understandable, and Generalizable the validity of the experiment is threatened. To address this concern, we compared crowdworker and expert ratings. The authors rated a sample of 50 questions under the multidimensional condition to generate a set of expert ratings. Expert ratings and crowdworker ratings were moderately correlated with a Pearson correlation of 0.62 (p<0.001). This suggests that the crowd was able to generate feedback comparable to experts and representative of the criteria (see Table 4).

*Perceptions of Feedback*
To test H1 and H2, we examine the effects of the different feedback conditions on perceptions of the feedback. Participants answered four Likert scale measures from 1 (Strongly Disagree) to 5 (Strongly Agree):

Q1: I agree with the feedback.
Q2: I found the feedback helpful.
Q3: I will change my answers in the future, based on the feedback.
Q4: It will be easy to incorporate the feedback into my future answers.

Participants answered these measures after viewing feedback for each of their five answers. For preliminary examination, we used participants as the unit of analysis, taking the mean over all five pieces of feedback. Table 5 reports the Likert measures for each rating by condition.

| | No-criteria | Unidimensional | Multidimensional |
|---|---|---|---|
| Q1 | 4.37±0.90 | 3.91±1.02 | 4.08±0.75 |
| Q2 | 4.42±0.73 | 3.83±1.12 | 3.92±0.89 |
| Q3 | 3.67±1.02 | 3.55±1.29 | 3.38±1.20 |
| Q4 | 4.41±0.78 | 3.79±1.21 | 3.78±1.10 |

**Table 5: Mean ± standard deviation for Likert measures Q1, Q2, Q3 and Q4, by condition.**

Overall, participants thought the feedback in general was helpful. We split Likert responses with 'agree' and 'strongly agree' as the positive category, and 'neutral', 'disagree', and 'strongly disagree' as negative. 78% of participants indicated agreement with the feedback (Q1). 75.1% found it helpful (Q2) and 70.3% found it easy to use (Q4). A smaller proportion of participants, 57.5%, said they would change their answers in the future (Q3). Participants in the no-criteria condition answered higher across all four measures compared to participants in the experimental conditions.

| | Rating | Unidimensional | Multidimensional |
|---|---|---|---|
| Q1 | 3.47* | 0.50 | 0.41* |
| Q2 | 2.32* | 0.28* | 0.27* |
| Q3 | 0.91 | 0.89 | 0.98 |
| Q4 | 1.29 | 0.37* | 0.30* |

**Table 6: Logistic regression results. Odds ratio effects are shown for Likert measures Q1-Q4. *p<0.05**

Table 6 shows the results of our logistic regression. Participants were .41 times as likely to agree (Q1) with the feedback in the multidimensional condition in comparison with the no-criteria condition (p=0.03). They were .28 and .27 times as likely to find the feedback helpful (Q2) under the unidimensional (p=0.002) and multidimensional (p=0.001) conditions. And they were .37 and .30 times as likely to say the feedback was easy to incorporate (Q4) in the unidimensional (p=0.01) and multidimensional (p<0.001) conditions. This contradicts H1—while we expected the inclusion of criteria to result in higher perceptions of utility, the participants found the criteria conditions less useful. Furthermore, our analysis did not find any support for H2. There was no significant difference between Likert measures for unidimensional and multidimensional conditions. The rating was a significant predictor of Likert measures— participants were 3.47 times as likely (p<0.001) to agree (Q1) and 2.32 times as likely (p<0.001) to find the feedback helpful (Q2) per star received.

*Effects of Feedback on Answer Quality*
We further tested behavioral measures of H1 and H2 by examining whether receiving the feedback had an effect on answer quality – as assessed through existing community metrics of answer ratings, 'thanks,' and 'Brainliest answer' indicators. 22 participants (7 no-criteria, 6 unidimensional, 9 multidimensional) out of 55 who viewed the feedback remained active, answering at least 5 additional questions.

| Condition | Rating | Thanks | Brainliest |
|---|---|---|---|
| No-criteria | -0.95±2.10 | -2.14±5.34 | -0.42±0.79 |
| Unidimensional | -0.24±2.85 | -3.16±6.21 | -0.17±1.83 |
| Multidimensional | -0.90±1.79 | 0.11±6.68 | 0.11±0.93 |

**Table 7: Mean ± standard deviation for pre-post changes in average ratings, 'thanks,' and 'Brainliest answer' for participants in each condition.**

Table 7 reports mean changes in each outcome by condition. Our ANOVA analyses found no significant differences between conditions in pre-post differences for 'thanks,' (F(2,19)=0.57, p=0.58) 'Brainliest answer,' (F(2,19)=0.40, p=0.68) or ratings (F(2,19)=0.02, p=0.98).

*Summary of Quantitative Results*
To summarize, a sizable proportion of answers in our dataset did not receive any feedback. The vast majority of ratings from the Brainly community were 4 or 5 stars, while a small proportion received 1 star. While crowdworkers approached expert evaluations of the constructs in each criterion, their no-criteria ratings only modestly correlated with Brainly users' ratings. After viewing the feedback intervention, Brainly users generally agreed with the crowdworker feedback and found it helpful. However, they were significantly more likely to say so if they received a high rating, and they were much less likely to find the feedback helpful or easy to incorporate in the criteria conditions.

The results contradicted our expectations as outlined in H1 and H2. Criteria positions feedback in relation to explicit goals. We therefore expected the criteria-based conditions to produce more agreement and be perceived as more helpful. Furthermore, we expected per-criterion ratings to provide more direction to recipients, making the feedback easier to incorporate. Instead, participants indicated the criteria-based conditions were unhelpful and difficult to incorporate. Furthermore, there were no significant pre-post differences in community ratings, 'thanks,' or 'Brainliest answer' counts between conditions. The results of our experiment raise several important follow-up questions:

- Did Brainly users agree with our theory-driven criteria?
- Why did Brainly users fail to respond to our criteria-based feedback design?
- In what ways are Brainly users' existing community norms and practices influencing their perceptions of answer feedback?

In the next section, we describe participant interviews that explain our surprising experimental findings.

## INTERVIEW
### Interview Procedure
To gain qualitative insight into answerer needs and behavior and explain the results of our experiment, we reached out to all participants from our experiment. 17 responded and 15 (10 male and 5 female) attended an interview conducted over video chat or phone. Interviewees were contacted over email and offered a $20 gift card as compensation for their time. We asked them about their usage of the feedback system on Brainly as well as their opinion of the criteria and answer ratings from the experiment. Interviewees also revisited the experimental feedback webpage during the interview and gave their thoughts about each of the answer ratings.

### Interview Analysis
We transcribed the interviews and examined them line by line using a general inductive approach [18]. We developed codes for the interview data as needed throughout the process, and after an initial pass, we recoded the data with the complete set of codes. Finally, codes and their associated text were thematically grouped and translated by the researchers into the below results summary.

### Interview Results
*Answerers Agree with the Criteria*
Overall, interviewees' responses suggest their interpretation of quality aligns with the dimensions from the criteria used in our experiment: Appropriate, Understandable, and Generalizable. One said, "I couldn't have said it any better myself, that that would be the formula for a good answer" (P14). Their responses evidenced that the criteria were aligned with answerers' goals. "I would use them to know where to improve my answers. If understanding was low, I'd make sure my answer was clear and understandable" (P1). They place importance on ensuring the asker internalizes the answer and can solve similar problems. "I always try to give answer [sic] in a generalized way, so that he would be able to answer the future questions of similar type" (P10). A good answer is one that gives a problem-solving strategy.

> "You have to tell the answer in such a way that they should get the idea. So, like in the future, if they have the same question, they can answer by themselves. So the main purpose most importantly to answer is the method" (P8).

Although answerers agreed with the criteria in general, they rejected feedback in the criteria-based conditions. Next, we discuss contextual factors that conflicted with the ratings.

*Answerers are Asker Focused*
Our interviews revealed that the needs of the asker are paramount. Interviewees stated they use feedback to confirm that their answer reached the asker, was helpful, and that the asker internalizes the idea behind the answer. One interviewee stated: "the asker's need is what I care about. What he's asking, I would like to answer him exactly what he needs." (P10). As a result of this asker focus, cues from the asker direct answerers when to elaborate.

> "Sometimes the question has the words, 'Please help,' or 'I can't get my head around this,' or something like that. Something that indicates that they are struggling with the question. In such cases, I make sure that I give them an explanation because they are really looking to learn" (P10).

This introduces a dilemma when the expectations of the asker conflict with the criteria. Some askers are not interested in explanatory answers. P8 explained, "if you put much effort, and put much theory, they don't like this. They just like the answer, which is to the point answer." Participants were acutely aware of times that asker's expectations did not match the criteria: "[the rater] wants us to do more than is necessary to answer the question, so we'll get a low score in generalizable because we just answered the question how [the asker] wanted, not a detailed answer" (P6). Thus, answerers' agreement with the criteria was conditional on the needs conveyed by the particular asker.

The prioritization of askers might also explain why a few answerers took issue with ratings from crowdworkers. They did not view crowdworkers as a reliable source of feedback because they do not know the asker's needs. "You should be aware of what you are rating. But I think most of Mechanical Turkers are not . . . [A one-star rating] doesn't mean the asker doesn't get the idea, or didn't get what he's looking for" (P4). The answerer's interpretation of quality lies in the asker's assessment, rather than a more objective external measure.

The asker focus may be reinforced by a lack of information about other viewers who may benefit from a high-quality answer. Brainly provides a searchable archive that is not immediately obvious from the answerer perspective, and answerers that don't ask questions might not be aware of it.

> "If I type an answer, is that just a one-to-one conversation with the people asking the question, or are other students looking at, reviewing all those answers to learn something? I have no clue if that's happening. So, does Brainly become like a Wikipedia of questions already answered, or is it just something that only happens in the here-and-now?" (P15).

Indeed, a few participants said they would like to be more aware of other viewers. P14 elaborated, "like on YouTube when you see a video, you see the views, and maybe you could do that for [Brainly]. That would be pretty interesting."

*Answerers Perceive Feedback as Binary*

The way answerers currently view feedback on Brainly also illuminates their reactions to our feedback. Community members use feedback on Brainly for two primary functions: providing affirmation when the answer is accurate and helpful, or providing critical feedback when the answer is unsatisfactory. 'Thanks' and ratings, as currently used on Brainly, are perceived as affirming and helpful for verifying the correctness of answers. "If it's correct, they would just five-star and give a 'thanks.' Which tells me that I substantially helped them and gave them the right answer to the question." (P11). Likewise, 4- or 5-star answer ratings from our experiment provided reassurance to answerers. Reflecting on a 4-star rating, one participant stated, "I didn't go into detail but I would give it the rating of four out of five . . . I wouldn't change it, as long as it's correct" (P11).

A second function of the feedback is to provide a cue to action to participants, indicating that a specific change is necessary to their answers. While affirmative feedback motivates answerers, several stated that the most helpful feedback comes when they are wrong.

> "Sometimes people say 'That's not the answer, this is the answer.' At that point, they comment on it. I would look, if I get such a comment, then it would be helpful for me. I would reevaluate my answer and if it's wrong, then it's helpful for me because I just corrected myself" (P10).

This affected participants' perception of the ratings from the multidimensional condition, which at times left participants wondering what they were supposed to change about their answers. One user suggested incorporating comments with feedback from the multidimensional condition.

> "You can only get so much out of stars. You don't know what they're thinking, actually. Like every little thought, maybe they could if there was a text box or underneath each of the three factors that correlated. If you wanted you could expand" (P6).

**DISCUSSION**

Providing answerers with feedback is a promising way to enhance the quality of answers and support self-regulated learning within social Q&A. We presented a study exploring criteria-based crowdworker feedback on Brainly.com, finding that answerers reacted negatively to criteria-based evaluation compared to the no-criteria condition. One limitation of this study is selection bias—only sufficiently motivated answerers viewed the feedback and completed the study. It is possible that users not represented in our results do not want to be evaluated, which may affect their response to criteria-based feedback as well. A better understanding of students' motivation to receive feedback in informal settings would be beneficial for designing feedback affordances.

Another concern with our approach is whether or not crowds provide sufficiently accurate feedback. In support of Luther et al. [12], we found that aggregated crowdworker ratings were moderately correlated with expert ratings. That U.S. Mechanical Turk workers can accurately assess middle school to college level content makes intuitive sense given their relatively high education levels [13]. Albeit some variation, high-quality answers received high ratings, and unmet criteria received low ratings from experts and crowdworkers alike. This indicates that a panel of three crowdworkers can provide a sufficiently accurate assessment of the criteria without any training. We might achieve a high correlation, a close star-for-star match between experts and crowdworkers, by careful selection or by training raters. The much weaker correlation between crowdworkers and Brainly ratings, however, suggests crowdworkers do not represent the Brainly community—they are outsiders.

Our results contradicted theoretical predictions about the use of feedback. Kluger and DeNisi state that explicit goals provide a standard for comparison, and feedback recipients are motivated to respond when their performance does not meet that standard. Instead, we found that the inclusion of criteria did not significantly affect recipients' willingness to change their answers and was perceived to be less helpful. Hattie and Timperley [8] suggest that feedback is more effective when it indicates to recipients where to go next. The increased specificity of one rating per criterion could direct recipients to an area for improvement. However, we found no significant differences between perceptions of feedback when multiple ratings were included.

While we hypothesized that the criteria-based feedback would be perceived more positively and be more effective for improving answers, the participants found the no-criteria feedback more helpful. We also found no differences in answerers' answer ratings, 'thanks,' or 'Brainliest answer' counts after the experiment. In scientific communities, such negative results are often brushed aside. But these results beg the question: why did answerers reject feedback that aligned with their goals and was objectively accurate?

Answerers have multiple reasons for participating: learning and helping others are two primary motivations. From the interviews, we found that answerers were focused on supporting askers, and unaware of their impact on other users viewing the archive. Kluger and DeNisi [9] suggest that the presence of multiple standards may result in a weighted overall evaluation of the feedback. We find that in social Q&A, the asker introduces a standard and answerers place more weight on the asker's standard. Because they were motivated to address the asker's need, answerers valued affirmation and requests for clarification directly from the asker. They dismissed external feedback despite accurate assessment. Their agreement with the criteria itself was situational, qualified by their focus on the asker.

Relying on askers to set the standard for answers may ultimately inhibit the efficacy of Q&A as a learning platform. Gazan [10] separates askers into 'seekers'—those who interact with the community and engage in conversation about their questions, and 'sloths'—those who post their homework verbatim and interact no further. Interview participants were aware of these differing uses of the site, and changed their standard for a good answer based on whether the asker was looking for elaboration. Dissuading answerers from elaborating may inhibit domain learning for everyone involved, as helpers learn when they generate explanations [14] and recipients learn as they internalize and apply answers [19], not by copying them verbatim.

Learning is one of the most commonly reported motivations for asking and answering questions on Brainly [4], and there exists great potential for promoting learning during the provision of peer help [14,19]. However, a number of issues need to be addressed for our crowdsourced, criteria-based feedback intervention to be effective. First, answerers are unaware of how writing explanatory answers could affect their own learning, and primarily look to corrective feedback as the primary learning mechanism on Brainly. Second, objectively valid external feedback is not perceived as a legitimate evaluation of the asker's needs. And finally, answerers are unaware of potentially large numbers of viewers turning up their answers while searching the web and looking for quality answers [15,16]. In order to promote learning during Q&A information exchange, we need to design a system that balances the priorities of satisfying askers and providing quality help.

## DESIGN IMPLICATIONS

**The benefits of explanatory answers to all community members should be salient to answerers.** Our findings revealed that for answerers, meeting askers' expectations trumped outsider feedback. While the criteria we outlined was agreed with in principle, for the goals to be adopted, community norms would need to change. These norms are embedded in the design of Brainly's interface—for instance, only the asker can mark the "Brainliest" answer, while a potentially large number of users [15] never leave a trace when they view the answer. Thus, viewers don't help to set the standard for quality answers. In addition, there are no affordances available to answerers to track their learning.

Designs must balance the goals of askers, answerers, and viewers. This could involve increasing the social presence of viewers as well as directing answerers' attention to the beneficial effect of composing answers on their own learning. Making answerers aware of how their answers impact themselves and others could encourage them to consider more than the asker's goals.

**Designs should raise the credibility of feedback givers.** Answerers distrusted the assessment of crowdworkers as it was not reflective of the asker's need. Answerers did not always recognize the utility of raters in providing an outsider assessment. Criteria-based, aggregated crowd feedback approached expert feedback in both our study and prior work [12,20]. Designs should establish the legitimacy of raters. Even in cases where the feedback giver is not an expert, the answerer should be given enough knowledge to make a credibility assessment about the rating. This trust is crucial for the effectiveness of outsider feedback.

**Feedback should be affirming and actionable.** Our interviews found that affirmation from others motivated them to continue participating. High ratings, 'thanks' clicks and especially positive comments helped to meet this need. This part of the current design on Brainly is working well. Other social Q&A sites seeking to increase participation should foreground user behaviors that affirm contributions. The findings of our interview also show that critique is highly valued, especially when it is specific and actionable. Participants looked to corrective comments on Brainly for learning about their incorrect answers, often stating that this is the most helpful kind of feedback. While comments, both affirmative and corrective, were perceived to be most valuable, this is also the type of feedback with the least presence, occurring on about 1 in 5 answers. Encouraging these kinds comments when users leave feedback could prove highly effective for increasing answerer contribution.

## CONCLUSION

Social Q&A offers a promising space for students from multiple communities to learn from each other. Effective answers have the potential to support learning for askers, answerers, and viewers. We tested a feedback intervention designed to evaluate the effectiveness of peer help, promote explanation, and ultimately incorporate learning into the information exchange. Based on our findings, we derived a set of design principles for how feedback should be offered to answerers in the future. Our investigation did reveal a need for feedback—these are opportunities to inform answerers about the effectiveness of their answers and direct them to improve. Designs will need to balance theories of feedback with the practical needs of the community to support learning processes that occur during the course of social Q&A.

## ACKNOWLEDGEMENTS

We offer our heartfelt thanks to the participants, folks at Brainly, and reviewers at Learning@Scale.